\documentclass[prl,noshowpacs,noshowkeys,twocolumn]{revtex4}
\usepackage{graphicx}
\usepackage{bm}
\usepackage{textcomp}
\usepackage{color}
\usepackage{amsfonts}

\begin{document}

\title{Colloidal topological insulators}
\author{Johannes Loehr$^{a}$, Daniel de las Heras$^{a}$, 
	Adam Jarosz$^b$, Maciej Urbaniak$^b$, Feliks Stobiecki$^b$, 
	Andreea Tomita$^c$, Rico Huhnstock$^c$, Iris Koch$^c$, Arno Ehresmann$^c$, Dennis Holzinger$^c$, and Thomas M. Fischer$^{a}$}\email{thomas.fischer@uni-bayreuth.de}
\affiliation{$^{a}$ Institute of Physics and Mathematics, Universit\"at Bayreuth, D-95440 Bayreuth (Germany).
	$^b$ Institute of Molecular Physics, Polish Academy of Sciences, 
	ul. M. Smoluchowskiego 17, 60-179 Pozna\'n (Poland).
	$^c$ Institute of Physics and Centre for Interdisciplinary Nanostructure Science and Technology (CINSaT), Universit\"at Kassel, Heinrich-Plett-Strasse 40, D-34132 Kassel (Germany)}
\date{\today}
\begin{abstract}{
	Topological insulators insulate in the bulk but exhibit robust
	conducting edge states protected by the topology of the bulk material.
	Here, we design a colloidal topological insulator and demonstrate
	experimentally the occurrence of edge states in a classical particle system. 
	Magnetic colloidal particles travel along the edge of two distinct magnetic lattices.
	We drive the colloids with a uniform external
	magnetic field that performs a topologically non-trivial modulation loop.
	The loop induces closed orbits in the bulk of the magnetic lattices.
	At the edge, where both lattices merge, the colloids perform skipping orbits
	trajectories and hence edge-transport. We also observe
	paramagnetic and diamagnetic colloids moving in opposite directions along the edge between two inverted
	patterns; the analogue of a quantum spin Hall effect in topological insulators.
	We present a new, robust, and versatile way of transporting colloidal
	particles, enabling new pathways towards lab on a chip applications.
}\end{abstract}

\maketitle

Topologically protected quantum edge states arise from the non trivial topology (non-vanishing Chern number)
of the bulk band structure~\cite{Hasan}. If the Fermi energy is located in the gap of the bulk band structure, like in an ordinary
insulator, edge currents might propagate along the edges of the bulk material. The edge currents are
protected as long as perturbations to the system do not cause a band gap closure. The topological mechanism at
work is not limited to quantum systems but has been shown to work equally well for classical photonic~\cite{Rechtsman,Perczel},
phononic~\cite{Kane,Mao}, solitonic~\cite{Paulose}, gyroscopic~\cite{Nash}, coupled pendulums~\cite{Huber},
and stochastic~\cite{Murugan} waves. It is also known that the topological properties survive the particle limit
when the particle size is small compared to the width of the edge.
In the semi-classical picture of the quantum Hall effect, the magnetic field enforces the electrons to perform closed
cyclotron orbits in the bulk of the material. Near the edge, the electrons can only perform skipping orbits, i.e.
open trajectories that allow electronic transport along the edge~\cite{Beenakker}. Considerable effort to simulate such
semi-classical trajectories has been undertaken~\cite{Beenakker,Davies,Shi,Montambaux}. Their experimental observation, however,
is quite difficult. So far, skipping orbits were only observed in two dimensional electron gas driven by microwaves \cite{Zhirov} and with
neutral atomic fermions in synthetic Hall ribbons \cite{Mancini}.

We present here the experimental observation in a colloidal system of skipping orbits and hence edge states.
These edge states allow for a robust transport of colloids along the edges and also the corners of the underlying
magnetic lattice. 

\begin{figure*}
\includegraphics[width=1.9\columnwidth]{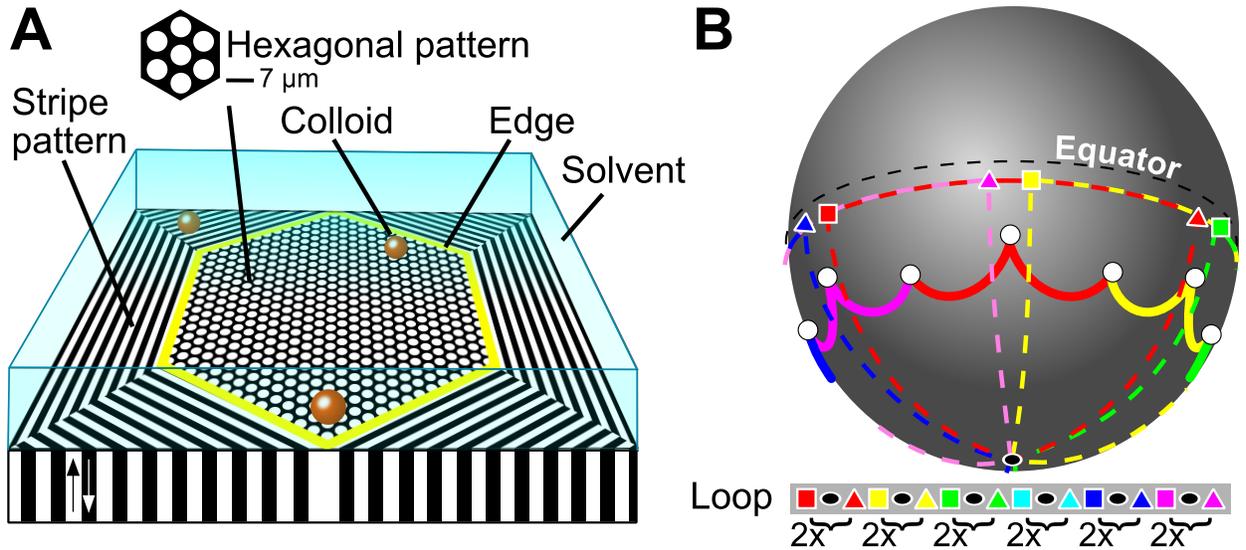}
\caption{{\bf Colloidal topological insulator}. ({\bf A}) Paramagnetic colloids are confined at a constant distance above a magnetically structured film of thickness $3.5$ nm with regions
of positive (white) and negative (black) magnetization perpendicular to the film. The film is a hexagonal lattice embedded into a stripe pattern.
({\bf B}) Control space $\cal C$ for a hexagonal lattice: twelve bifurcation points (empty circles) connected by segments (solid lines). Winding around
a pair of equal color segments moves the colloids one unit cell along one of the hexagonal directions. The modulation loop used here is indicated by dashed
lines and also in the legend. The loop starts at the red square and winds anticlockwise twice around each pair of equal color segments.} 
\label{fig1}
\end{figure*}

We use a thin cobalt-based magnetic film lithographically patterned via ion bombardment~\cite{CBF1998,KET2010,tp3}. The pattern consists of a patch of hexagonally arranged circular domains (lattice constant $a\approx 7\,\mu\textrm{m}$) surrounded by a stripe pattern, see Fig.~\ref{fig1}A. Both magnetic regions consist of alternating domains magnetized in the
$\pm z$-direction normal to the film. Paramagnetic colloidal particles of diameter
$2.8\,\mu\textrm{m}$ are immersed in water (or aqueous ferrofluid) and move at a fixed
elevation above the pattern.  Hence, the particles move in a two-dimensional
plane that we refer to as action space, $\cal A$. A uniform time-dependent
external magnetic field $\mathbf{H}_{\text{ext}}(t)$ of constant magnitude ($H_{\text{ext}}=4\,\textrm{kA}/\textrm{m}$)
is superimposed to the non-uniform and time-independent magnetic field of the pattern $\mathbf{H}_{\text{p}}$.
We vary $\mathbf{H}_{\text{ext}}(t)$ on the surface of
a sphere that we call the control space $\cal C$ (Fig.~\ref{fig1}B).

\begin{figure*}
\includegraphics[width=1.9\columnwidth]{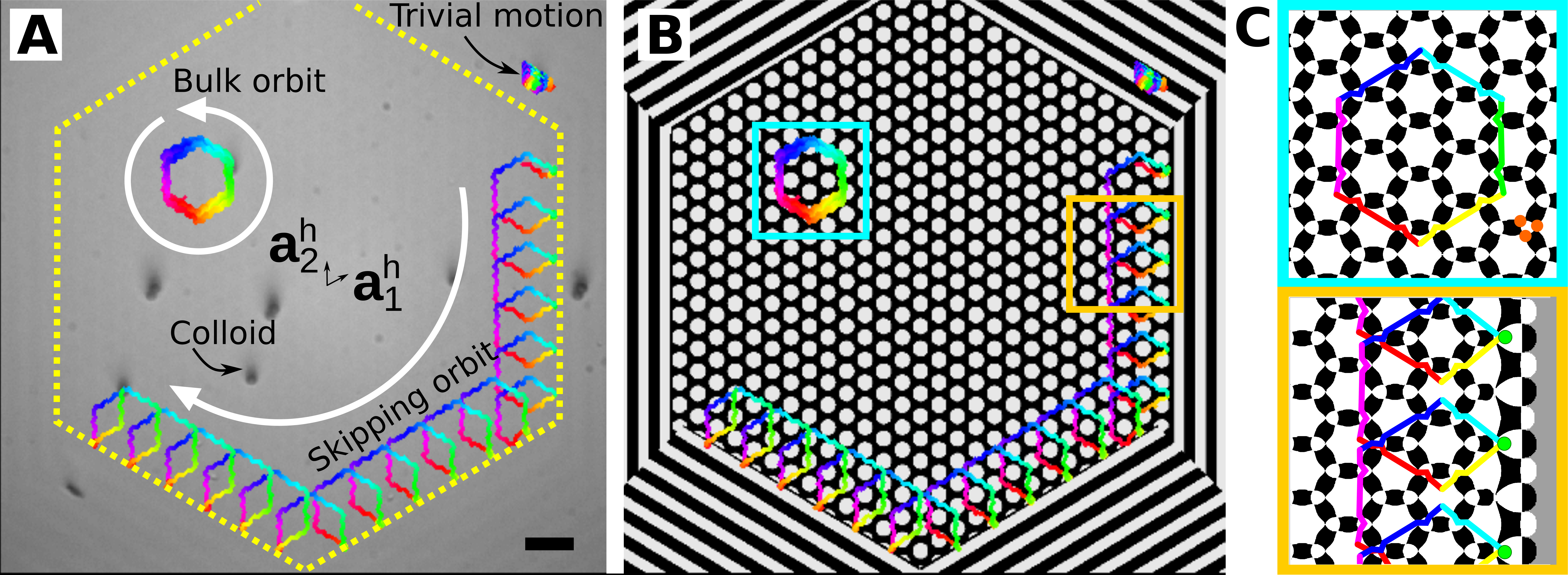}
\caption{{\bf Skipping and bulk orbits}. ({\bf A}) Microscopy images of paramagnetic colloidal particles at the end of a transport process. Scale bar $15\,\mu\text{m}$.
Three trajectories are shown and indicated: a bulk orbit on top of the hexagonal pattern, a skipping orbit at
the edge, and the trivial motion of a particle above the stripe pattern. The color of the trajectories matches that of the modulation loop (Fig.~\ref{fig1}B).
The edge between the hexagonal and the stripe patterns is indicated by a dashed yellow line. 
The colloidal motion can be seen in \cite{MovieCite}.
{\bf (B)} Colloidal trajectories superimposed on the actual magnetic pattern. {\bf (C)}
Bulk (top) and skipping (bottom) orbits superimposed on action space, ${\cal A}$. The white (black) areas are allowed (forbidden) regions for the colloids. The grey
area is the indifferent region on top of the stripe pattern. In the skipping orbits the colloids skip the green section of the bulk-trajectory. Three gates connecting 
adjacent allowed regions are highlighted with orange circles in the top panel.}
\label{fig2}
\end{figure*}

In Refs.~\cite{tp1,tp2,tp3} we demonstrate how bulk transport of colloids above different magnetic lattices
can be topologically protected. For each lattice symmetry there exist special modulation loops of 
$\mathbf{H}_{\text{ext}}$ in $\cal C$ that induce transport of colloids in $\cal A$. These loops share a 
common feature, they wind around special objects in $\cal C$~\cite{tp1,tp2,tp3}.
For a hexagonal, six-fold symmetric lattice, the control space of paramagnetic colloids is characterized
by twelve points connected by six pairs of segments (Fig.~\ref{fig1}B). A modulation
loop encircling one pair of segments in $\cal C$ transports the colloids in $\cal A$ one unit cell along 
one of the six possible directions of the hexagonal lattice
$\pm\mathbf {a}_1^{\text{h}},\pm\mathbf {a}_2^{\text{h}},\pm(\mathbf {a}_1^{\text{h}}-\mathbf {a}_2^{\text{h}})$, with 
$\mathbf{a}_1^{\text{h}}$, and $\mathbf{a}_2^{\text{h}}$ the lattice vectors. E.g.,
encircling the red segments in Fig.~\ref{fig1}B transports the paramagnetic 
particles into the $(\mathbf {a}_1^{\text{h}}-\mathbf {a}_2^{\text{h}})$-direction. If we now rotate the
modulation loop such that it encircles the yellow segments, the transport occurs along $\mathbf {a}_1^{\text{h}}$, i.e.
the transport direction rotates $\pi/3$ with respect to the previous $(\mathbf {a}_1^{\text{h}}-\mathbf {a}_2^{\text{h}})$-direction.
Using these results we construct a hexagonal cyclotron orbit of hexagonal side length $na^{\text{h}}$,
$n=1,2,3...$ of a paramagnetic particle above the bulk of a hexagonal lattice. The corresponding cyclotron modulation
loop in $\cal C$ consists of six connected parts. Each of the six parts of the loop
winds $n$ times around a different pair of segments of $\cal C$, and therefore transports a particle $n$ unit cells
along one of the hexagonal directions in ${\cal A}$. In Fig.~\ref{fig1}B we show a $n=2$ cyclotron modulation loop in $\cal C$.
The corresponding colloidal trajectory in $\cal A$ is depicted in Fig.~\ref{fig2}A,B.
Colloidal particles above the bulk of the hexagonal pattern perform closed hexagonal
cyclotron orbits of the desired side length. The cyclotron modulation loop does not wind around the
special objects in $\cal C$ for a stripe pattern (located on the equator of ${\cal C}$~\cite{tp3}).
Hence, the colloids above the stripes are not transported. For the current modulation loops, the edge between the hexagonal and the stripe pattern
is an edge between topologically non-trivial and trivial patterns, and allows for the existence of edge states, as it is the case in quantum topological
insulators. Edge states are possible for those particles close to the edge between
both patterns. The paramagnetic particles perform skipping orbits, see Fig.~\ref{fig2}A,B and movie S1 \cite{MovieCite}. That is,
the particles do not follow all six directions of the closed cyclotron orbit but skip one of the hexagonal directions.
The skipping direction is different for different orientations of the edge.
A successive series of these skipping orbits results in an open trajectory along the edge direction.
The skipping orbits allow for robust transport in armchair edges, i.e. edges oriented along one of the six directions
of the hexagonal lattice, and also around the corners where two edges join.

The skipping of single directions can be explained by taking a closer look at the bulk transport mechanism,
explained in detail in Refs.~\cite{tp1,tp2,tp3}. Here, we just summarize the main concepts.
For a fixed external field there are stationary points of the magnetic potential in each unit cell of the magnetic lattice. The 
colloids are transported by following the minima of the magnetic potential. 
Action space can be split into allowed, forbidden, and indifferent regions.   
In the allowed regions the stationary points are minima, whereas in the forbidden regions the only possible stationary
points are saddle points. No extrema of the total magnetic potential exist in the indifferent regions, which are present only in stripe patterns.
Therefore, the colloids can occupy only allowed and indifferent regions. We show in Fig.~\ref{fig2}C the split of ${\cal A}$
into the different types of regions for the current pattern. Two adjacent allowed regions touch each other at special 
points in ${\cal A}$ that we refer to as the gates. To achieve transport between adjacent unit cells, a colloid has
to pass through two gates. In the bulk of the hexagonal lattice transport is possible along all
six crystallographic directions~\cite{tp1}. However, close to the edge this is no longer true. The necessary gates to
transport the particle into the edge or parallel to it are no longer available. In consequence those particles close to the
edge have to perform skipping orbits. 

\begin{figure}
\includegraphics[width=1\columnwidth]{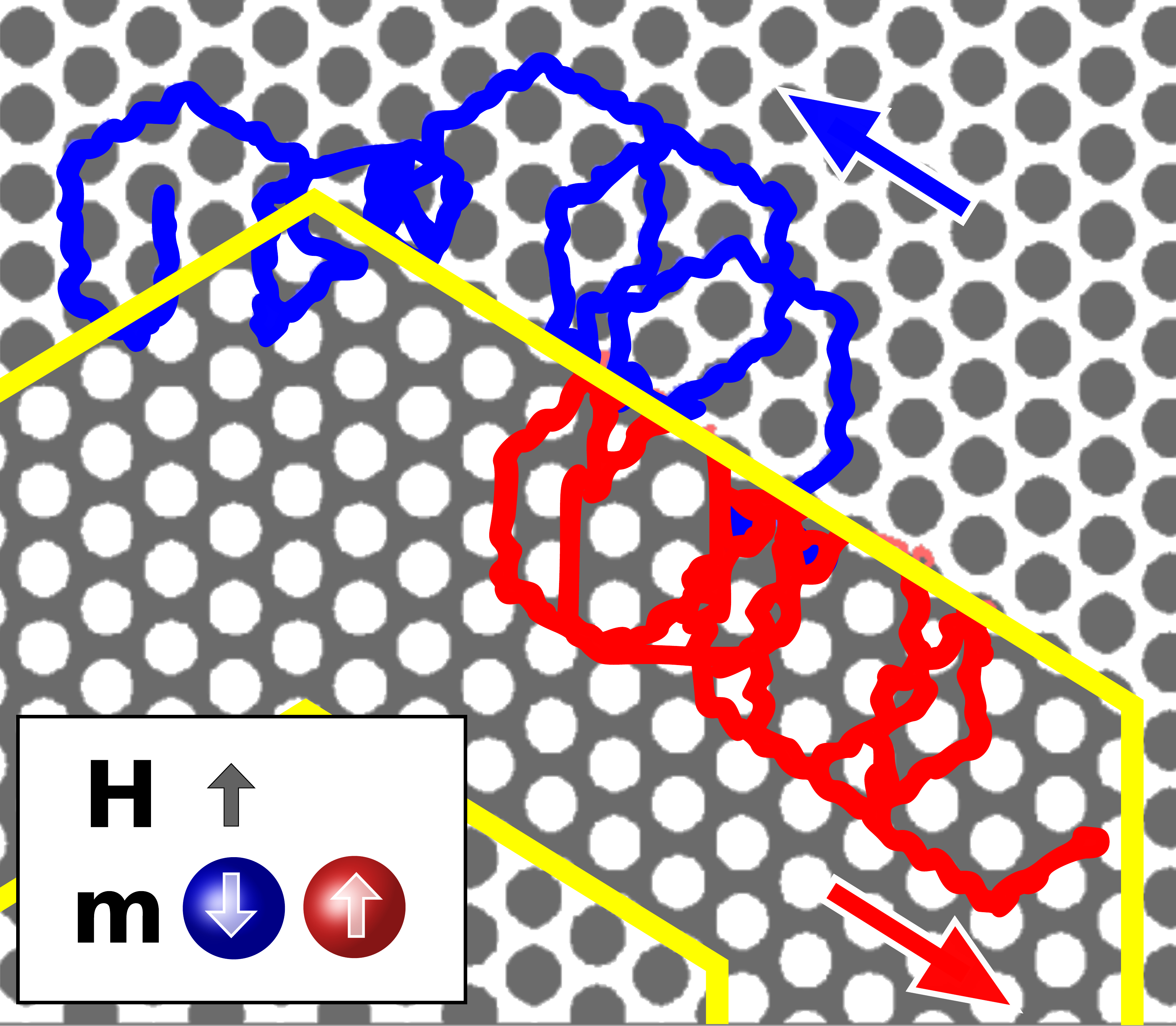}
\caption{{\bf Colloidal spin Hall effect}. Trajectories followed by a paramagnetic (red) and a diamagnetic (blue) particle along
the edge (yellow lines) of two hexagonal magnetic patterns with opposite magnetization. The particles are driven by the modulation loop
shown in Fig.~\ref{fig1}B. The magnetic moment ${\mathbf{m}}$ of the paramagnets (diamagnets) is parallel (antiparallel) to the total
magnetic field ${\mathbf{H}}$. A video of the colloidal motion is provided in \cite{MovieCite}.}
\label{fig3}
\end{figure}
 
Transport above an infinite lattice remains unchanged if the magnetization of both the pattern and the colloidal particles is inverted.
Thus, diamagnetic particles (magnetic holes) on a pattern respond to the external field in the same way as paramagnetic particles
on an inverted pattern. In Fig.~\ref{fig3} and movie S2 \cite{MovieCite} we show the chiral response of paramagnetic
and diamagnetic particles at the edge between two hexagonal patterns with inverted magnetization.
The susceptibility of the ferrofluid-based solvent is set to a value
in between the effective particle susceptibilities.
We use the same cyclotron modulation loop as before. The loop induces anticlockwise cyclotron orbits of paramagnets
on the bulk of one lattice and of diamagnets on the bulk of the inverted lattice.
Near the edge, both types of colloids perform skipping orbits above the pattern that is non-trivial for them.
Since the edge between both patterns is located in opposite directions from the center of the corresponding orbits, 
the skipping directions are antiparallel. Hence, this results in skipping orbits along the edge where
both types of particles move on opposite sides of the edge in opposite directions. This represents the colloidal analogue 
of the quantum spin Hall effect~\cite{Bernevig}, in which electrons of opposite spins move in different directions along the same edge.
 
In this work we experimentally demonstrated how to realize a colloidal topological insulator. Like in the semi-classical picture of the quantum Hall effect,
particles above the bulk of the material move following closed orbits, and particles close to the edge perform skipping orbits giving rise to robust edge states.
The fine details of the pattern are irrelevant to determine the bulk transport properties~\cite{tp3}. 
At the edges, however, transport does depend on the details of the pattern. E.g., the exact positions of the stripes with respect to the circular domains and the orientation 
of the edges influence the edge states. Similar effects are also observed in graphene where e.g., only certain edges support edge states.
The versatility of our robust colloidal transport opens the possibility to transport multiple particles along multiple different edges into different
directions using just a unique external modulation. Colloids can be used to mimic aspects of molecules~\cite{Glotzer} and atoms~\cite{poon}.
Our colloidal topological insulator goes a step further and mimics the behaviour of electrons in a colloidal system.

\section*{Acknowledgements}
A. T. is supported by PhD fellowship of the University of Kassel.

 \emph{}


\begin{thebibliography}{10}
\bibitem{MovieCite} A movie S1 of the coexistence of cyclotron orbits with skipping orbits at the edge
 			and trivial orbits in the stripe pattern and a movie S2 of a paramagnetic particle and a diamagnetic particle at the edge traveling in opposite directions can be viewed free of charge in XXXXXXX.
\end{thebibliography}
\end{document}